\documentclass[amsmath,amssymb,twocolumn,superscriptaddress]{revtex4-2}

\usepackage[utf8]{inputenc}
\usepackage{color}
\usepackage{graphicx}
\usepackage{dcolumn}
\usepackage{times}
\usepackage{soul}
\usepackage{siunitx}
\usepackage{afterpage}
\usepackage{xcolor}
\usepackage{tabularx}
\usepackage{placeins}
\usepackage{bm}
\usepackage{setspace}
\usepackage{siunitx}
\usepackage{soul}

\usepackage{hyperref}
\usepackage{epstopdf}

\begin{document}

\title{Cooperative effect of local active stresses on the macroscopic contractility of elastic fiber networks}

\author{Abhinav Kumar}
\affiliation{Department of Physics, University of California, Merced, Merced, CA 95343, USA}
\affiliation{Cluster of Excellence `Physics of Life', TU Dresden, Dresden 01307, Germany}

\author{David A. Quint}
\affiliation{Lawrence Livermore National Laboratory, Livermore, California 94550, USA}

\author{Kinjal Dasbiswas}
\email{kdasbiswas@ucmerced.edu}
\affiliation{Department of Physics, University of California, Merced, Merced, CA 95343, USA}

\begin{abstract}
The collective action of actively contractile units embedded in elastic biopolymer networks plays a crucial role in regulating the network's macroscopic mechanical response. Here, we investigate how the macroscopic boundary stress in model elastic fiber networks depends on the number and nature of embedded contractile units, each exerting an isotropic force dipole, as well as on the bending stiffness of fibers. We find that the macroscopic stress increases nonlinearly with the number of dipoles due to mutual stiffening of initially soft, bending-dominated networks. Using effective medium theory, we relate this enhanced contractility to an increase in the effective average network coordination number due to constraints imposed by the force dipoles. By comparing three distinct force dipole models that differ in their local structures, we demonstrate that the specific manner in which an active unit constrains the network strongly influences the onset and nature of the stiffening transition. Our results highlight that not only the quantity but also the local geometry of force-generating units critically determines the macroscopic mechanical behavior. This framework provides a physical basis for understanding how biological systems—such as molecular motors in the cytoskeleton, or adherent cells in the extracellular matrix—can modulate network-scale nonlinear elastic properties through local tuning of active force-generating units.

\end{abstract}

\maketitle

\renewcommand*\rmdefault{bch}\normalfont\upshape
\rmfamily
\section*{}
\vspace{-1cm}

%%%FOOTNOTES%%%
\footnotetext{\textit{$^{a}$~Cluster of Excellence `Physics of Life', TU Dresden, Dresden 01307, Germany.}}
\footnotetext{\textit{$^{b}$~Lawrence Livermore National Laboratory, Livermore, California 94550, USA.}}
\footnotetext{\textit{$^{c}$~Department of Physics, University of California, Merced, Merced, CA 95343, USA; E-mail: kdasbiswas@ucmerced.edu}}
\footnotetext{\dag~Electronic Supplementary Information (ESI) available: }
%%%END OF FOOTNOTES%%%

\section{Introduction}

Animal cells use mechanical forces generated in their actomyosin cytoskeleton to change shape \cite{lecuit_11}, divide \cite{green_12}, and move  \cite{gardel_10, gardel_15}.  These mechanics-driven processes are essential for biological functions such as tissue morphogenesis and wound healing \cite{barriga2018}, as well as for tumor progression \cite{discher2017matrix}. 
These mechanical forces are generated by molecular motors of the myosin family by transducing ATP-driven chemical reactions into mechanical work  \cite{howard}. Specifically, myosin motors bind to and slide actin filaments of opposite polarity to produce force distributions that deform and contract the surrounding cytoskeletal network. Cells use these forces to deform and restructure their extracellular medium \cite{abhilash_14}, as well as to probe and sense its mechanical properties \cite{Notbohm2015}. The latter provides a pathway for cell-cell mechanical communication \cite{schwarz2013physics, Tang11, noerr2023optimal} in addition to chemical signaling.  The ability of cells to strongly contract their surrounding medium is particularly important for biological functions such as wound healing \cite{doha2022disorder},cardiomyocyte beating \cite{Nitsan2016}, and clot stabilization \cite{kim2017quantitative}. 

Both the extracellular matrix (ECM) and cytoskeleton (CSK) are biopolymer networks. They typically occur as hierarchical structures, where individual filaments are bundled by crosslinks into fibers, which in turn entangle to form networks \cite{burla2019mechanical}. While details in their structure vary, their short-time response to internal active forces is dominated by non-affine and heterogeneous deformations  characteristic of disordered fibrous networks \cite{HeussingerPhysRevLett2006, PicuSM2011,broedersz_14, feng2016nonlinear}. These softer bending and buckling modes lead to unusual elastic response not expected in linear elastic materials, such as rigidity transition under external or internal shear \cite{onck2005alternative, sharma2016strain, broedersz_11, doi:10.1073/pnas.1815436116} as well as uniaxial strain \cite{prachaseree2025towards},  negative normal stress \cite{janmey2007negative}, and buckling-induced softening \cite{conti2009cross, zakharov2024clots}, resulting in a renormalization of their Poisson ratio \cite{malakar2025rectification}. When poised near a rigidity transition threshold, such networks display large strain fluctuations \cite{Wyart2008Elasticity} , sensitive response to perturbations \cite{broedersz2011criticality}, and force amplification \cite{xu_15, Ronceray16}, which are all desirable for biological function. Apart from their biological relevance, disordered elastic networks are potentially applicable in designing desired response in synthetic metamaterials, including tunable elastic moduli \cite{Reid2018}, topological edge modes \cite{zhou2018topological}, memory storage \cite{keim2019memory}, and physical learning \cite{stern2023learning}. 

Fibrous networks are minimally modeled as a depleted lattice of springs with both bending and stretching stiffness, where disorder is introduced by randomly removing bonds \cite{das2012redundancy, broedersz2011criticality, Ronceray16, kumar2023range, broedersz_14,  shivers2019scaling, ArzashPhysRevE2022, Mao2022, zakharov2024clots, majumdar2025non}. The probability of bonds being present, $p$, is tuned to reach a target average coordination number per node, $6 p$  for a triangular lattice in 2D. The macroscopic mechanical response of such an elastic network to applied forces, depends on the single fiber mechanics, as well as the network geometry, particularly its connectivity represented by the average coordination number. Importantly, such networks undergo rigidity percolation transition as the coordination number is increased \cite{broedersz_14}:  first from floppy to a rigid phase which resists shear at $p_{b} \simeq 0.45$, and then from a bending- to stretching- dominated response at  $p^T_{CF} = 2/3$. This latter is the isostatic point corresponding to Maxwell's constraint counting argument in 2D  for a triangular lattice \cite{maxwell1864calculation}, where the number of bond constraints balances the degrees of freedom of a network node. The elastic response of such disordered elastic fiber networks to internal forces, actively generated by cells and motors, has been considered in prior models \cite{broedersz_11, Sheinman2012ActivelyStressed, Ronceray16, goren2020elastic}. However, how the contractile response depends on the bending stiffness of fibers and configurations of localized active stresses has not been systematically investigated. While rigidity transitions and associated critical exponents can depend on network dimensionality \cite{broedersz2011criticality, Chen2024FieldTheory, Zhang2025strainstiffening}, here we use a 2D elastic network model to minimally demonstrate how mutual stiffening by active force units may arise. We do not aim to characterize in detail here the rigidity phase transition from the bending- to stretching-dominated regime with increasing number of active units.

\textit{In vitro} experiments with cells cultured in ECM-like gels \cite{doha2022disorder, fernandez2009compaction} as well as reconstituted acto-myosin networks \cite{ideses2018spontaneous} show large scale contraction due to active forces generated by cells or myosin motors, respectively. Of note, Ref. \cite{doha2022disorder} shows that the macroscopic gel compaction exhibits a sharp transition dependent on cell density. At low cell density, the network barely contracts, while above a critical cell density, the cells are able to coordinate their forces to achieve large contraction.  Here, by modeling the contractility of cells or motors as active, force-producing units embedded in an elastic fiber network, we aim to show how such cooperative effects may arise. We are also motivated in part by our previous works which showed that easy fiber bending reduces the range of force transmission away from the localized force dipole \cite{kumar2023range, zakharov2024clots}. However, the role of individual fiber bending mechanics and network connectivity in determining macroscopic network contraction was not quantified in detail. Previous studies have shown that  contractile force dipoles stiffen a fiber network by ``pulling out’’ floppy modes \cite{broedersz_11, Chen2011}. Here, we ask: how many additional constraints does each force dipole effectively impose?

In this work, we measure macroscopic contractility of under-coordinated fiber networks as a function of the density of active force-producing units as well as of the bending stiffness of individual slender fibers. We use an established effective medium theory to quantitatively identify the approximate number of constraints imposed by force dipoles from their measured stiffening of the network. Our work shows that even without mechano-chemical feedback processes whereby cells actively regulate their force production, the nonlinear mechanical properties of fiber networks allow for cooperative effects and sensitive tuning of macroscopic network contractility. We also show that the local structural details of how the contractile forces, applied by cells or molecular motors, are important in determining the macroscopic mechanical response of the fiber network.

\section{Model}

\begin{figure*}[ht]
    \centering
    \includegraphics[width=18cm]{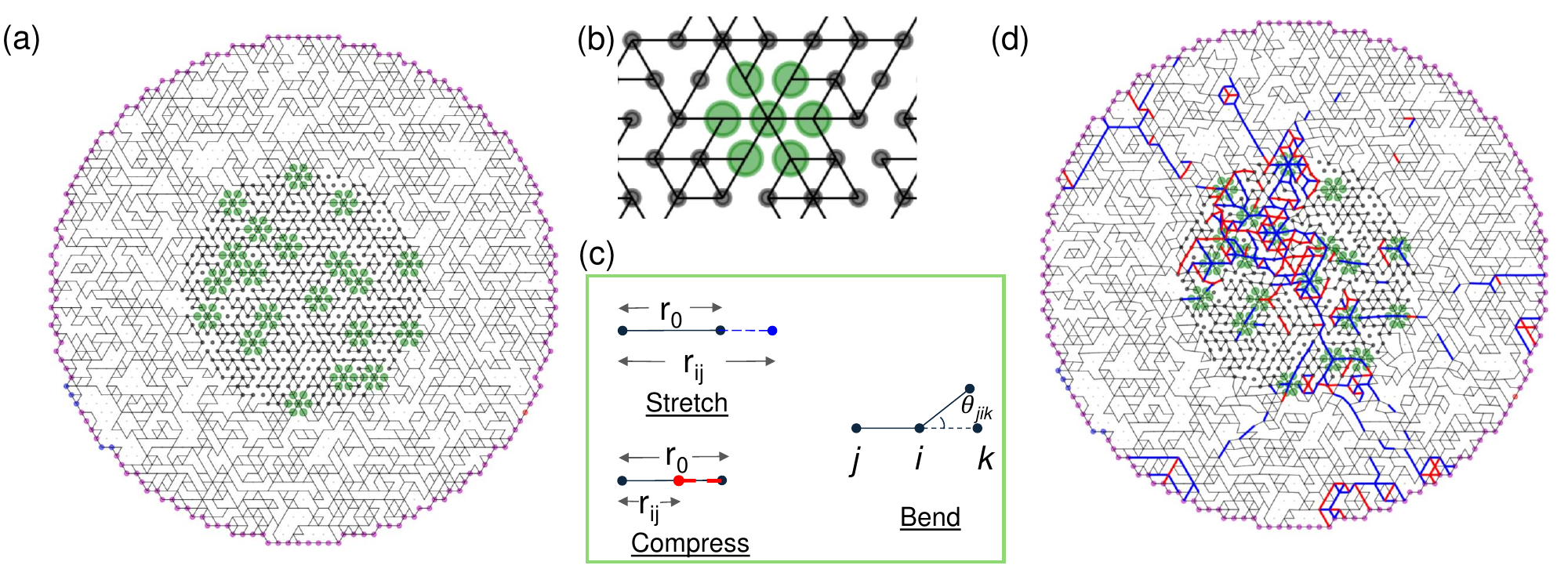}
    \caption{ \textbf{Model setup of elastic fiber network with embedded force dipoles.}
    (a) A representative network configuration with a circular outer boundary ($R_{2} = 25$), with fixed boundary nodes (colored in magenta). Isotropic force-producing units (nodes colored in green) are randomly placed in an inner circular region ($R_{1} = 12$), marked with darker colored nodes. For each dipole number, a total of 10-40 random dipole placements are generated in the inner region.  
    The network configuration in the annular outer region ($R_{1}  <  r < R_{2}$), is varied over 10-25 different realizations for each random dipole placement. 
    (b) A representative force-producing unit is depicted. It comprises a central node and six outer nodes, all shown in green. All six radial bonds coming out of the central node are required to be present during network construction. These bonds are given a rest-length of 0.9, less than the initial length of all bonds, which is set to 1.0. Therefore, these bonds are in a state of prestress and tend to contract towards their rest lengths as the network reaches its force balanced mechanical equilibrium configuration. The net effect is to produce an isotropic, contractile, local force dipole on the network.
    (c) The individual bonds in the network can stretch and compress, and are modeled as linear springs. Additionally, co-linear bonds can also bend by changing their relative angle with  a bending energy cost, typically much smaller than the stretching/compression energy cost.
    (d) The final network configuration at mechanical equilibrium, \textit{i.e.} after the network shown in (a) is relaxed to its energy minimum state using the conjugate gradient method. Here,  highly stretched(compressed) bonds are colored red(blue) for visualization purposes, if the magnitude of bond strain exceeds a threshold value, $\epsilon_0 = 10^{-6}$.  
    }
    \label{fig:Model_contraction}
\end{figure*}

We design an elastic network model to capture the macroscopic contractility of biological networks in response to internal motor-generated stresses. We deform the network with localized sources of contractile, isotropic stress. These local active units are seeded within a prescribed region of the elastic network, while we measure the forces transmitted to the network boundary. 

To capture the inherently disordered nature of biopolymer networks, we use a randomly depleted triangular lattice of springs, a commonly employed model for these systems \cite{das2012redundancy, broedersz_14, Ronceray16, broedersz2011criticality, kumar2023range}. Each bond connecting neighboring nodes in the network acts as a Hookean (central force) spring which resists stretching and compression with stiffness constant, $\mu$. To model fiber bending, we include angular springs between each pair of collinear bonds. These resist deviations in the relative angle between these two bonds from collinearity, with a bending stiffness constant, $\kappa$. 
The total elastic energy of the network is given by,
\begin{equation} 
E_t = \frac{\mu}{2 r_{0}}\sum_{\langle ij \rangle}  (r_{ij}-\bar{r}_{ij})^2 + \frac{\kappa}{r_{0}} \sum_{\langle jik \rangle} {2\sin^2(\theta_{jik}/2)}, 
\label{eq:total_energy}
\end{equation}
where, $\mu$ and $\kappa$ are the stretching and bending stiffness moduli, $\bar{r}_{ij}$ is the rest length of a bond connecting the $i^{th}$ and $j^{th}$ nodes, $r_0$ is the initial length of each spring, set equal to the lattice spacing of the undeformed triangular lattice, and $\theta_{jik}$ is the angle between bonds connecting nodes $j$, $i$ and $k$, respectively. In our depleted lattice model, the first term corresponding to stretching is present only if nodes $i$ and $j$ are connected by a bond.  Similarly, the second term corresponding to bending is present only if both nodes $i$ and $j$ as well as $i$ and  $k$ are connected.  We do not consider nonlinear elastic effects in the constitutive relationship for single fibers, such as stiffening under stretch and fiber buckling due to compressive forces in the present model. To avoid combining multiple coupled effects, we focus on the effects of transverse bending alone. Practically, as well, the buckling threshold may not be attained for thicker fibers, such as networks of bundled actin filaments \cite{bernheim_18} that are under smaller forces. 

Such a bond-depleted network then models a fiber network, where each fiber corresponds to a continuous set of collinear bonds. The length of the fibers in the model is then related to $p$, the probability of bonds being present. This also represents the mean coordination number of the nodes in the network, through  $\langle z \rangle =6p$. Biological fiber networks are thought to have a mean coordination number between $3 < \langle z \rangle < 4$, corresponding to fibers branching or crossing over, respectively \cite{Gardel04,sharma2016strain, wyse2022structural}. In the depleted triangular lattice, these correspond to bond probability values in the range, $p=0.5 -0.67$. In this work, we set $p=0.55$ as a representative value of coordination for biopolymer networks, to demonstrate the dependence of network contractility on fiber bending stiffness and applied stress. This choice ensures that our networks are not too close to the rigidity transition thresholds, where large fluctuations occur in network mechanical response, while remaining in the under-coordinated, bending-dominated regime, that is expected to be biologically relevant. 

The simulated domain is chosen to be a circular region containing a triangular lattice of springs, as shown in Fig.~\ref{fig:Model_contraction}. We choose a circular geometry since we are interested in macroscopic contractility at the network boundary, that is expected to be isotropic on average. This contractility is generated by a spatially random distribution of isotropic, active, force-generating units embedded within the network. Biologically, this corresponds to a nearly circular cell shape or isotropic aggregates of myosin motor filaments, as occurring in asters. The active force-producing units are randomly placed within an inner circular region of radius $R_{1}$, while nodes at the outer boundary of radius $R_{2}$ are held fixed to facilitate measurement of the boundary forces. For results reported in the main text, the outer radius is $R_{2} = 25$ and the inner radius is $R_{1} = 12$. We also consider a larger size in Appendix A, where we show that our main conclusions are robust to system size variations.

The contractile active stresses, generated by motors in the cytoskeleton or cells in the ECM, is modeled using a distribution of localized and isotropic, contractile active units. 
To realize the isotropic deformation by an active unit, here a contractile hexagon (marked by green nodes in Fig.~\ref{fig:Model_contraction}), we reduce equally the rest-length of all six bonds connected to the central node of the active unit, $\bar{r}_{dip} = 0.9 r_0$. The rest-length of all passive elastic springs in the network is set to $r_{0} = 1$, the initial undeformed length of all bonds. We note that with this choice, equal and opposite forces cancel out, resulting in a net zero force monopole, as must be the case for forces internal to a mechanical medium. Each such active unit does produce a dipole moment of forces, and will thus henceforth be referred to as an ``isotropic force dipole'' \cite{ben2015response}.  In general, molecular motor-generated active stresses will also have anisotropic components \cite{schwarz2013physics}, which we leave out here for simplicity and focus on isotropic contractility. We note that active deformations may be applied in an elastic network in a variety of ways \cite{goriely2017five}, either as a fixed ``active force'', or a fixed ``active strain''. Here, we choose the latter. While this choice is guided by practical numeric convenience, we note that for the bond-diluted networks we consider, this choice corresponds to fixed displacement of dipole nodes. This is in fact consistent with observations of cells on soft elastic substrates. As part of their \emph{mechanical homeostasis}, cells do in fact induce fixed displacement on soft substrates \cite{feld2020cellular}, while they maintain fixed stress on stiffer substrates \cite{ghibaudo2008traction}.  
 
While in the rest of the network, bonds are randomly removed to meet a certain $p$ value for the whole network, the dipole has a central node that is always kept fully coordinated, see Fig.~\ref{fig:Model_contraction}b. This choice helps to efficiently transmit the contractile force to the surrounding network. 
The six outer nodes of the force dipole are disconnected from each other but are always connected to the central node. In other words, bonds radial to the central node of each force dipole are always present, and the bonds transverse to the radial direction that connect the outer dipole nodes are removed. This modeling strategy (termed ``Model 1'') allows the outer dipole nodes to move inward easily (since the transverse bonds are removed), thereby enhancing the deformation of the surrounding network. In subsequent sections, we explore two other modeling choices to show how the details of local force application have a significant effect on macroscopic contractility.

This inner region ($r < R_1$) that contains the force dipoles can itself be considered a macroscopic dipole that exerts stress on the surrounding annular region \cite{ben2015response}.  This emergent stress arises from the interaction of  multiple dipoles. Each force dipole ``sees'' a different local environment that is influenced by the forces and constraints imposed by the other dipoles. The net contractility or macroscopic dipole moment is thus expected to depend on fiber stiffness parameters, local connectivity, as well as dipole strength and distribution.

We randomly generated $10$ dipole positions in the inner circular network for a fixed number of dipoles ($N_d$). Additionally, we generated $10$ random realizations, per dipole configuration, of the outer annular network bounded by $R_1$ and $R_2$, shown in Fig. ~\ref{fig:Model_contraction}. We then average the results from these dipole and network realizations to compute the quantities of interest that follow in this manuscript.
Additionally, in the SI,  we expanded the number of network-dipole configurations simulated for a representative set of dipole numbers ($N_d = 1, 5, 10, 20, 40$), to $1000$.  We realized $1000$ independent simulations by expanding the number of dipole configurations to $40$ and outer network configurations to $25$. In the SI Fig. S10, we show that the mean far field dipole, $\langle D_{far} \rangle$, is indeed captured well by $100$ independent simulations to within $95 \%$ confidence intervals of the $1000$ simulations.  For the single force dipole ($N_d=1$) case in particular, we found the results were strongly sensitive to specific network-dipole configuration, and we present averages over $1000$ simulations in the main text.

In each simulation, we minimize the elastic energy in Eq.~\ref{eq:total_energy} numerically using the conjugate gradient method to obtain the force-balanced or mechanical equilibrium configuration of the network. Further details on the computational procedure and parameter choices are discussed in Appendix B.

\section{Results}
\subsection{Macroscopic boundary stress increases nonlinearly with number of contractile units}

We first aim to characterize how the net macroscopic contractility of a bond-diluted fiber network depends on the applied local deformations.  Fiber network models with bonds diluted below the isostatic point ($p < p_{cf}$) are expected to undergo large bending-dominated deformations in response to applied shear, whether external or like in our system, internal. Since these soft deformation modes allow fibers to bend without appreciably stretching, we expect that the bonds will be under lower tension. Thus, only a part of the locally applied forces will be transmitted to the boundary of the network. This intuition is visualized in SI, Fig. S1 where boundary force vectors are plotted for a fully connected ($p=1$) and a depleted ($p=0.55$) network. To quantify the force transmission to the boundary, we measure the stress at the boundary. Specifically, we follow previous works \cite{ronceray2015connecting, Ronceray16} in  calculating a boundary dipole tensor, or the ``far field'' dipole moment, $D_{far}$, defined to be the trace of the dipole moment of the forces measured at the fixed outer boundary nodes,

\begin{equation} \label{eq:dfar}
D_{far} = \sum_{\substack{ i \\ 
                          i \in boundary}}
                \mathbf{f_{i}} \cdot \mathbf{r_{i}}, 
\end{equation}
where $\mathbf{f}_{i}$ is the force measured at the $i^{th}$ node on the boundary, and $\mathbf{r}_{i}$ is the position vector of this node, here measured from the origin of coordinates located at the center of the circular network domain. Through the virial theorem, this quantity is related to the isotropic, contractile stress measured at the boundary. Intuitively, this  corresponds to the amount of force transmitted from local force dipoles to the system boundary through the elastic network.  This quantity serves as our measure of the macroscopic network contractility induced by a configuration of local dipoles. 

Next, we systematically investigate how this macroscopic force transmission depends on the mechanical properties of individual fibers and the number density of the active force-producing units, for a network with given average coordination, $p$.  This network response function of interest may be formally expressed as $\partial \langle D_{far} \rangle/\partial \phi$. Here, the angular brackets indicate an average over network and dipole configurations, and $\phi$ is a non-dimensional density of force dipoles, defined in Appendix A.    

\begin{figure*}[ht]
    \centering
    \includegraphics[width=18cm]{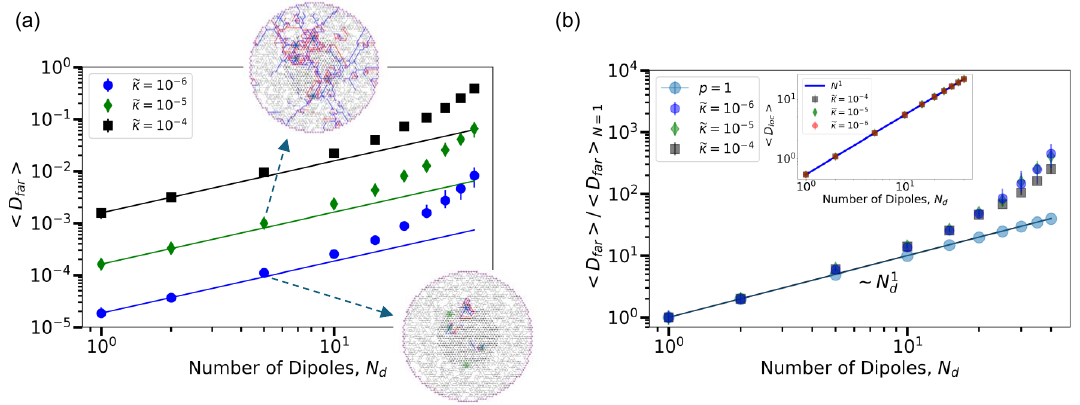}
    \caption{\textbf{Scaling of average far-field force dipole moment with bending modulus and number of force dipoles in bending-dominated networks with $\boldsymbol{p=0.55}$.} 
    (a) The average far-field dipole moment, $\langle D_{far} \rangle$, a measure of force transmission to the network boundary, increases linearly with dipole number $N_d$ for low $N_d$, but shows positive (upward) curvature at higher $N_d$.
    The straight lines show a linear scaling of $D_{far}$ with $N_d$.
    $\langle D_{far} \rangle$  increases approximately linearly with the reduced fiber bending modulus $\widetilde{\kappa}$, but the onset of nonlinearity is at a similar  value,  $N_d \gtrapprox 15$, for all three representative $\widetilde{\kappa} \ll 1$ cases simulated. The insets show two representative simulated network configurations with $N_d = 5$ dipoles each, but different bending moduli: $\widetilde\kappa = 10^{-5}$ (top left) and $\widetilde\kappa  = 10^{-6}$ (bottom right). Blue (red) colors indicate highly stretched (compressed) bonds above a strain magnitude threshold, $\epsilon_0 = 10^{-6}$. The softer network ($\widetilde\kappa = 10^{-6}$; bottom right) has visibly fewer highly tensed or compressed bonds (``force chains'') than the stiffer network ($\widetilde\kappa = 10^{-5}$; top left).  
    (b) Average Far-field dipole moments, $\langle D_{far} \rangle $, scaled by the corresponding average value for one dipole unit $N_{d} = 1$, at the same $p$ and $\widetilde\kappa$ values. $\langle D_{far} \rangle $ values for bond-diluted ($p=0.55$) networks at different bending moduli, $\widetilde\kappa$, collapse on the same master curve, indicating universal nonlinear stiffening of bending-dominated networks with increasing dipole number. The $p=1$ networks, by contrast, show linear increase of $D_{far}$ with $N_d$, characteristic of linear elastic media. [Inset] Local dipole moment scales linearly with the number of dipoles, for all networks with $p=0.55$ with different bending moduli. Thus, our simulation procedure ensures an approximately identical local force dipole is exerted by each active unit.
    Each plotted value of $\langle D_{far} \rangle$ is an average over all bending-dominated networks obtained from a total of $100$ simulations ($10$ dipole configurations $\times 10$ network configurations), except for the one dipole case ($N_d=1$), where  averaging is over 1000 simulations. Error bars indicate 95 $\%$ confidence intervals on the mean value obtained from bootstrapping.
    }
    \label{fig:dfar_nd}
\end{figure*}

The results of measured network ``far-field dipole moment'', $\langle D_{far} \rangle$, each averaged over all bending-dominated networks, are shown in Fig.~\ref{fig:dfar_nd}. For this analysis, we exclude the few cases of stretching-dominated networks that arise at higher dipole number ($N_d = 35, \, 40$). We choose to remain in the bending-dominated regime, where network bending energy is lower than stretching energy. We make this choice to avoid large fluctuations near the transition from bending to stretching-dominated regime, that skew the distribution of  $D_{far}$ values. In Fig. ~\ref{fig:dfar_nd}a, we show the trends in $\langle D_{far} \rangle$ vs. number of dipoles, $N_{d}$, for
three different values of the ratio of bending to stretching stiffness parameters, $\widetilde{\kappa}$. These could correspond to different values of fiber thickness or inter-fibril bundling in the case of composite fibers in biopolymer networks \cite{kumar2023range, zakharov2024clots}. 
In diluted $p=0.55$ networks, $\langle D_{far} \rangle$ increases linearly with the number of dipoles for low $N_d$. This suggests that the material remains linearly elastic. However, for $N_d \geq 15$, the slopes of all three curves steepen (Fig. ~\ref{fig:dfar_nd}a), suggesting stiffening of the medium at increased dipole density. In the SI Fig. S2, we show the corresponding result for $\langle D_{far} \rangle$ based on all network simulations carried out, including the stretching-dominated cases. As expected, the enhanced stretching further steepens the nonlinear dependence of $\langle D_{far} \rangle$ on $N_d$.

In contrast to the dependence on the dipole number, we find that $\langle D_{far} \rangle$ increases linearly with increasing bending modulus. The linear scaling vs. bending moduli is apparent from  Fig.~\ref{fig:dfar_nd}a and also shown explicitly in SI, Fig. S3. The dependence can be qualitatively understood from the representative network configurations shown in the insets to Fig. ~\ref{fig:dfar_nd}a. Here, bonds carrying strains of magnitude greater than a threshold value ($|\epsilon_0| = 10^{-6}$) are colored blue (extension) or red (compression). There are more colored bonds when there is more imposed stress in the network, as confirmed in  the SI, Fig. S4. The two networks shown as insets to Fig.~\ref{fig:dfar_nd}a are identical in configuration, and differ only in the  bending modulus of co-linear bonds. Clearly, the network with a relatively higher bending modulus ($\widetilde\kappa = 10^{-5}$) has more strained bonds, and a correspondingly higher $D_{far}$, compared to the network with $\widetilde\kappa = 10^{-6}$. The network with lower $\widetilde\kappa$ has fibers that bend more easily in response to applied shear, leading to less stretching/compression of its bonds. Thus, it also transmits less force to the boundary.  Quantitatively, a macroscopic elastic modulus of the network (say, the shear modulus, $G$) in the bending-dominated regime is determined by the fiber bending modulus, the only relevant stiffness or deformation energy scale \cite{broedersz_14}, leading to $D_{far} \sim G \sim \widetilde{\kappa}$.

By scaling the $\langle D_{far} \rangle$ values with their corresponding value for the single dipole network ($N_d=1$), we show in Fig.~\ref{fig:dfar_nd}b that the data for different bending moduli can be collapsed onto a single nonlinear master curve, suggesting a universal scaling across different $\widetilde\kappa$ values. The non-linear scaling seen in Fig. ~\ref{fig:dfar_nd}a at high $N_d$ values is also present in the scaled curves in Fig.~\ref{fig:dfar_nd}b. In contrast to the nonlinear data for the diluted networks, we show in Fig.~\ref{fig:dfar_nd}b that for $p=1$ networks, the $\langle D_{far} \rangle$ remains linear in $N_d$, even at higher $N_d$. 

We show in the inset of Fig. ~\ref{fig:dfar_nd}b that in contrast to $D_{far}$  the total local dipole moments, $D_{loc}$,  scale linearly with $N_d$ for all bending moduli, $\widetilde{\kappa}$. The definition of the local dipole moment of active forces is provided in Appendix C together with the procedure of calculating it from our simulation. The linear scaling of $D_{loc}$ with $N_d$ establishes that in our model, each force dipole applies active forces locally independent of other dipoles. Biological cells in elastic media may adapt their contractility to the local strain, that may be generated by nearby cells \cite{schwarz2013physics}. Such substrate-mediated cell-cell interactions could be an additional source of non-linearity in the net contractility of the cell-substrate system. However, our results in Fig.~\ref{fig:dfar_nd} demonstrate that such non-linear, co-operative effects can exist even when the contractility of each cell is identical and independent of each other.  The origin of the nonlinearity here is the complex elastic response of a dilute, bending-dominated fiber network, as we now show.

In a linear elastic medium, corresponding here to a $p=1$ network, the total force transmitted to the boundary of the network is expected to be equal to the active force exerted on the network by the local force dipoles. Mechanical force balance leads to a mean stress theorem \cite{eshelby1956continuum, gurtin1973linear, carlsson2006contractile, ronceray2015connecting} that relates the local dipole moment ($D_{loc}$) to the far-field dipole moment ($D_{far}$) and the mean stress ($\bar\sigma$) over a finite elastic domain. See Appendix D for the statement of the theorem in Eq.~\ref{mean_stress} and its derivation for a continuous elastic medium. For a linear elastic medium with clamped boundary conditions, the mean stress vanishes, leading to a conservation of the dipole moment measured at the boundary with the dipole moment applied locally and internally within the network: $D_{far} = D_{loc}$. The total local dipole moment is just the sum of the dipole moments for each active unit, $D_{loc} = N_{d} D_{loc,1}$. From the dipole conservation theorem stated here and detailed in Appendix D, we expect $D_{far}$ to scale linearly with the number of dipoles, $N_d$, for the $p=1$ networks. This is indeed verified in Fig. ~\ref{fig:dfar_nd}b and in the SI, Figs. S5-S6.

\begin{figure*}[ht]
    \centering
    \includegraphics[width=18cm]{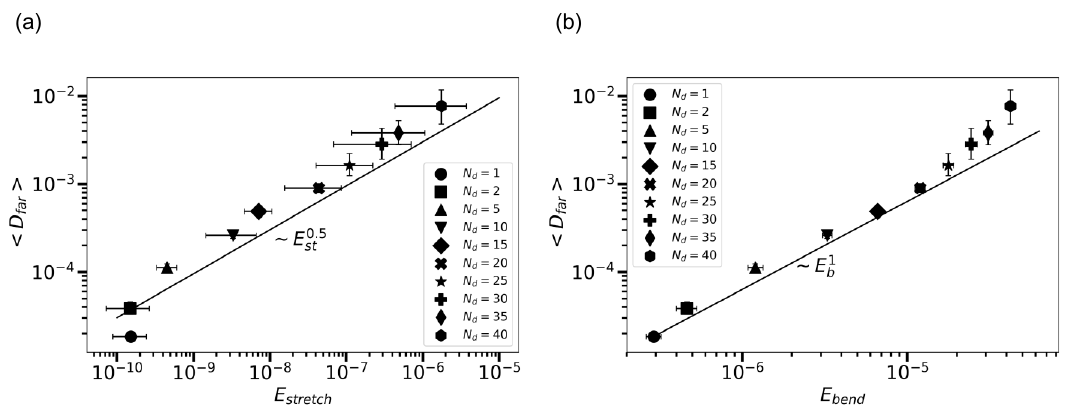}
    \caption{ \textbf{Scaling of $\boldsymbol{\langle D_{far} \rangle}$ with stretching and bending energies for bending- dominated networks at $\boldsymbol{p=0.55}$ and $\boldsymbol{\widetilde\kappa=10^{-6}}$.}
    (a) $\langle D_{far} \rangle $ values scale as the square root of the average stretching energy as dipole number increases. This is the expected stress-energy relationship from usual elasticity.
    (b) $\langle D_{far} \rangle $ scales linearly with the average bending energy as dipole number increases. This is a departure from the usual elastic stress-energy relationship, and can be explained based on how bending contributes to bond tension (see main text). Each plotted value of $\langle D_{far} \rangle$ is an average over all bending-dominated networks obtained from a total of $100$ simulations ($10$ dipole configurations $\times 10$ network configurations), except for the one dipole case, where  averaging is over 1000 simulations. Error bars indicate 95 $\%$ confidence intervals on the mean value obtained from bootstrapping.
    }
    \label{fig:dfar_en}
\end{figure*}

However, dipole conservation does not necessarily hold for nonlinear elastic media, such as the diluted $p=0.55$ networks considered here. In fact, $D_{far} \ll D_{loc}$ for the cases considered here, as seen from comparing Fig.~\ref{fig:dfar_nd}a with the inset in Fig.~\ref{fig:dfar_nd}b. SI Figs. S4 and S5 taken together show how dipole conservation is satisfied for $p=1$ networks, while it is strongly violated for sub-isostatic ($p<p_{CF}$), dilute networks. The nonlinear increase in $D_{far}$ with $N_{d}$ for dilute networks shows that adding one dipole to a multi-dipole system adds more to the boundary stress, $D_{far}$, than the previous contribution, per dipole. Intuitively, this can be understood as a consequence of the network being stiffened (attaining a higher macroscopic elastic modulus) due to the pre-stress exerted by the dipoles already present. A stiffer medium transmits more force to the boundary for the same applied local dipole. The effect of the prestress can also be understood as extra constraints induced by the force dipoles, which raise the effective connectivity, $p_{eff}$, of the network. We discuss in section 3.3 how these constraints can be estimated.

We also tested that the predicted nonlinear increase in macroscopic contractility with dipole number was robust to system size variations by simulating networks with twice the inner and outer radii compared to the original network ($R_1 = 24$ and $R_2 = 50$) . To compare the two networks, we define a dimensionless dipole packing fraction, $\phi_d$, detailed in Appendix A. We show in Fig. ~\ref{fig:dfar_L128} that the far-field dipole moments scale similarly with dipole density for different system sizes. In the next section, we quantify how much each force dipole stiffens the network. This leads to the non-linear increase in macroscopic contractility, measured by $D_{far}$, we found in this section.

\subsection{ Macroscopic network contractility scales differently with stretching and bending energy}

It is well-known that elastic fiber networks exhibit a crossover from bending to  stretching-dominated response, as the applied shear increases \cite{head_03}. We now characterize the partitioning of elastic energy between bending ($E_{bend}$) and stretching ($E_{st}$) deformation modes, as the \textit{internal shear} applied by the force dipoles increases.

In Fig.~\ref{fig:dfar_en}, we show the 
$\langle D_{\textrm{far}} \rangle$ values, averaged over all bending-dominated networks, at each dipole number, $N_{d}$, against the corresponding average elastic energies of these networks. Error bars indicate 95\% confidence interval obtained through bootstrapping 100 independent simulations with 10,000 resamples (SI Fig. S7). The $\langle D_{far} \rangle$ data is seen to collapse on distinct power law curves, showing definite but different scalings vs. stretching and bending energies.
The scaling with stretching energy is expected from elasticity theory. For Hookean springs, the elastic energy is quadratic in stress, giving $E_{st} \sim D_{far}^2$, which implies the observed scaling of $D_{far} \sim E^{1/2}_{st}$. 

\begin{figure*}[ht]
    \centering
    \includegraphics[width=18cm]{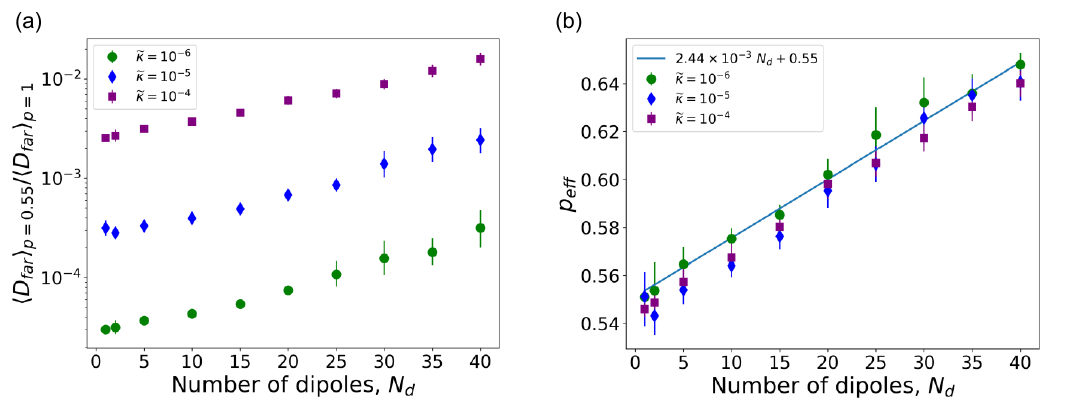}
    \caption{\textbf{Dipoles impose additional constraints and increase effective network connectivity in bending-dominated networks.}
    (a) Ratio of $\langle D_{far} \rangle$ for networks at $p = 0.55$ and $p = 1$. This is a measure of the fraction of forces applied by local dipoles, that is transmitted to the boundary, and therefore of effective network stiffness. This ratio increases with increasing dipole number and bending modulus.
    (b) The effective bond probability, $p_{eff}$, of a corresponding passive network with equivalent stiffness, $\mu_m$, to our local dipole-stressed networks, is obtained using effective medium theory (EMT) (see main text for details). The linear dependence of $p_{eff}$ on $N_d$ shows that each dipole applies a fixed number of constraints, corresponding to Eq. ~\ref{eq:peff} in the text.  The value of the slope of linear fit (shown here for $\widetilde\kappa$ = $10^{-6}$) corresponds to $n^{d}_{c} = 17.2 \pm 0.1$ constraints per dipole. 
    The increase in $p_{eff}$ with $N_d$ is however very similar for the three different  bending modulus $\widetilde\kappa \ll 1$ values simulated. This indicates that a similar number of floppy modes are ``pulled out'' for networks in this bending-dominated regime. Each plotted value of $\langle D_{far} \rangle$ is an average over all bending-dominated networks obtained from a total of $100$ simulations ($10$ dipole configurations $\times 10$ network configurations), except for the one dipole case, where  averaging is over 1000 simulations. Error bars indicate 95 $\%$ confidence intervals on the mean value obtained from bootstrapping.
    }
    \label{fig:peff}
\end{figure*}

On the other hand, bending deformations are non-affine and cannot be described within a continuum elastic framework. The scaling of bending energy can instead be simply understood by considering the basic unit of bending in the network: a single fiber represented by three nodes connected by two collinear bonds. When deflected by a small angle, $\delta \theta$, there is a small axial component of tension developed due to the bending force, given by $f_{b, \parallel} \sim f_{b} \cdot \delta \theta$. The bending force itself can be written based on the bending energy term in Eq.~\ref{eq:total_energy} and dimensional considerations, as $f_{b} \sim \kappa \cdot \delta \theta/r^{2}_{0}$. This term can be used to calculate a transverse spring constant for node displacement due to bending \cite{wilhelm_03,grill2021directed}. Combining these two relations shows that the axial tension developed as a result of bending is quadratic in the small angle deformation, and thus scales linearly with the bending energy: $f_{b, \parallel} \sim \kappa \cdot \delta\theta^{2}/r^{2}_{0} \sim E_{b} \cdot r_{0}^{-1}$. This axial component of force developed  in a bent fiber is in turn transmitted to neighboring fibers along its length, and contributes to stress developed at the boundary, measured by $D_{far}$.  This approximate argument based on a single fiber bending helps rationalize the linear scaling, $D_{far} \sim E_{bend}$, seen in simulations in Fig.~\ref{fig:dfar_en}b. We show also that this scaling is obeyed by networks with different bending moduli (SI, Fig. S8), as well as in results that include stretching-dominated networks (SI, Fig. S9). Overall, our analysis is consistent with the tension resulting from bending and stretching propagating differently through the heterogeneous network \cite{grill2021directed}.

The different scalings of stretching energy, $E_{st} \sim D_{far}^{2}$, and bending energy, $E_{bend} \sim D_{far}$ with boundary stress, suggest that these terms are in competition, and predict an expected transition from relatively more bending energy at low dipole number, to more stretching energy at higher dipole number. 
However, unlike the  macroscopic network bending-to-stretching rigidity transition obtained under external shear \cite{sharma2016strain, shivers2019scaling}, the dominant contributions to stretching and bending energy in our case are localized in the network. The localization of bending energy is visualized in SI Fig. S11, and may occur far from a dipole location. Additionally, SI Fig. S12 shows that changing $\widetilde\kappa$ scales the value of stored bending energies, but does not change their spatial localization.

\subsection{Constraints applied by force dipoles increase effective average coordination number of network}

It is intuitively apparent that internal force dipoles stiffen a sub-isostatic fiber network by ``pulling out'' its soft bending modes. We now compare our simulation results for macroscopic boundary stress with the predictions of \textit{effective medium} theory (EMT) \cite{feng1985effective, das_07, MaoPhysRevE2013, das2012redundancy, broedersz2011criticality}. This allows us to quantitatively assess the effective number of additional constraints created by force dipoles. The EMT approach  approximately captures the macroscopic elastic response of a disordered elastic network to applied stresses, by mapping to a homogeneous elastic network where each spring has modified stiffness, $\mu_m$. This effective stiffness depends on the bond coordination, $p$ and lattice type, with increased bond dilution resulting in lower $\mu_m$. While originally formulated for diluted spring lattices \cite{feng1985effective}, it was later extended to fiber networks with finite bending stiffness \cite{das_07, das2012redundancy, MaoPhysRevE2013, broedersz2011criticality}, the results of which are directly applicable to our model simulations. 
 
While previous works calculate the effective stiffness of passive, unstressed networks, it is not trivial to include active force dipoles, that exert pre-stress, in the theory.  We therefore adopt an indirect approach where we infer an effective  spring stiffness, $\mu_m^{\textrm{sim}}$, from the boundary forces measured in our simulations. We then utilize an established version of EMT for passive networks with bending constraints \cite{das_07, das2012redundancy}, to extract an effective bond probability, $p_{eff}$, for our dipole pre-stressed networks. This inferred network coordination is expected to be higher than the physical coordination of the network, since it includes the effect of the additional constraints induced by the dipoles, \emph{i.e.} $p_{eff} > p$.

The first step in our process for inferring $p_{eff}$ relies on the argument that the stress measured at the clamped boundary of an elastic medium scales with its elastic modulus. Consider for example, a one-dimensional series of springs, each of stiffness $k$, fixed at both ends. If one of the internal springs is contracted by a distance $\delta$, the force measured at the boundary is $ f \sim k \delta$. Similarly, for a continuous elastic medium, the boundary force, and therefore the far-field dipole moment, is expected to be linear in its elastic modulus. For the effective medium, that is, a fully connected network of springs, the continuum elastic modulus scales with the spring stiffness \cite{seung1988defects}, $ G \sim \mu_m$. 

Thus, we consider the factor by which the measured $D_{far}$ is reduced for the diluted ($p=0.55$) network in comparison to the $p=1$ network. We then obtain, 
\begin{equation} \label{eq:mu_m}
\frac{\mu^{\text{sim}}_{m}}{\mu} =  \frac{\langle D_{far,p=0.55} \rangle}{D_{far,p=1}}, 
\end{equation}
where $\mu_m^{sim}$ is the effective medium spring stiffness inferred from boundary force measurements in the simulation, whereas $\mu=1$ is the spring constant of each bond physically present in the actual network. Fig. ~\ref{fig:peff}a shows how this effective stiffness measured from boundary force attenuation, $\mu^{\text{sim}}_m$, increases nonlinearly with the number of dipoles, at the three different values of $\widetilde\kappa$ simulated. We further observe that the $D_{far}$ values measured for $p=0.55$ networks are much smaller than the corresponding $p=1$ networks. This is expected since highly diluted networks have fewer springs to transmit the local dipole forces to the boundary. More precisely, in sub-isostatic networks ($p<0.67$ for 2D triangular networks), the local stresses applied by force dipoles are primarily stored in the bending modes of the fibers. This reduces stretching of the fibers and therefore, force transmission to the boundary.

\begin{figure*}[ht!]
    \centering
    \includegraphics[width=18cm]{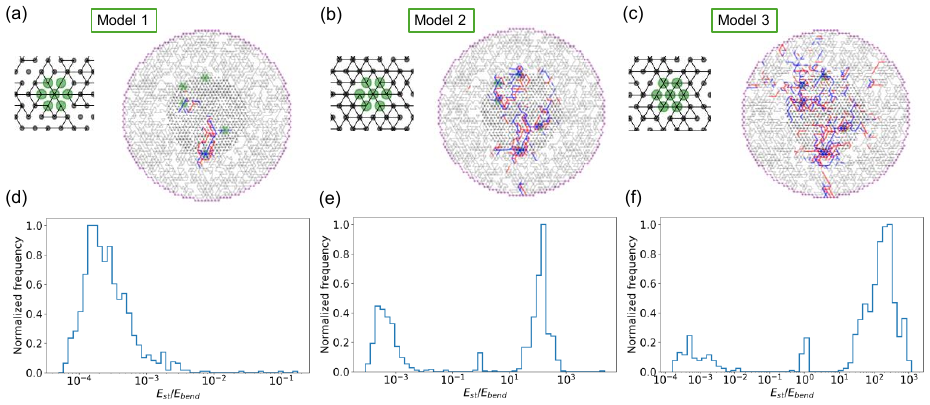}
    \caption{\textbf{Comparison of three local force dipole models showing significantly different distributions of network deformation energy.}
    In all three dipole models, the central dipole node is fully coordinated, i.e. all six radial bonds are present. Contraction is applied by reducing their rest lengths relative to initial value. 
    (a) Model 1: All six transverse bonds between six outer  dipole nodes are removed. Representative force-balanced network configuration shows very few strained bonds (red/blue).
    (b) Model 2: Transverse bonds connecting the outer dipole are present with a probability $p$. Representative force-balanced network configuration shows more strained bonds (red/blue) than model 1. 
    (c) Model 3: All six transverse bonds connecting the outer dipole nodes are present. Representative force-balanced network configuration shows more strained bonds (red/blue) than models 1 and 2. The distribution of ratios of network stretching and bending energies shown in (d,e,f) are obtained from  $1000$ representative simulations for a five-dipole configuration in $p=0.55$ networks with $\widetilde\kappa=10^{-6}$ ($40$ random dipole placements, and $25$ random outer network configurations for each dipole placement, leading to $1000$ total simulations). All three histograms (d,e,f) are normalized by maximum bin height.
    (d) Model 1: All networks simulated are bending-dominated ($E_{st}/E_{bend} < 1$), as expected for small deformations of sub-isostatic networks.
    (e) Model 2: A fraction of the random network realizations are stretching-dominated ($E_{st}/E_{bend} > 1$) .
    (f) Model 3: A majority of networks turn stretching-dominated despite being at $p=0.55$.  
    This suggests that progressively more pre-stress and thus effective constraints are applied by the force dipoles as we go from model 1 to 2 to 3.    
    }
    \label{fig:three_models}
\end{figure*}

Next, we use an established mapping between bond probability, $p$, and macroscopic network shear modulus, $G$, from an EMT calculation \cite{das_07, das2012redundancy}, to obtain the effective coordination, $p_{eff}$, for our pre-stressed networks with dipoles. This calculated theoretical relationship of $G \sim \mu_m$ with $p$ that we use as an intermediate step, is reproduced in SI, Fig. S13. 
We note that in this work no external stress is applied to the network and all the forces are internal to the system. The network design, featuring a circular domain with fixed boundary and internal force dipoles, allows us to calculate an effective stiffness of the network without imposing external shear. Our quantitative measurement of the effective stiffness from boundary forces, and the subsequent estimation of $p_{eff}$ from that using EMT, represents a key conceptual advance of this work. This central result is shown in Fig. ~\ref{fig:peff}b for the same family of bending-dominated networks presented in Fig. ~\ref{fig:dfar_nd}.

The linear trend in $p_{eff}$ vs. $N_d$ suggests that each dipole introduces a fixed number of constraints, or alternatively, removes a fixed number of floppy bending modes, at least for the regime studied here. We can estimate this number from the slope of the linear increase and Maxwellian constraint counting \cite{maxwell1864calculation}.

Generally, the number of constraints for an unstressed spring network of $N$ nodes is given by half the number of springs $N_{c} = zNp/2$, since each spring is shared between two nodes. Here, $z$ is the coordination number of each node of the undiluted network, being $z=6$ for a triangular network.  Let us now suppose that by applying restrictions on the rest length of dipole bonds, which creates pre-stress, we add an unknown  number of constraints per dipole, $n_c^d $. Therefore, we can write the enhanced number of constraints for a pre-stressed network with $N_d$ dipoles, as $N_{c} = zNp/2 + n_c^d N_d  $. We then can find the enhanced bond probability of an equivalent passive network without dipoles, 
\begin{equation}
    \frac{zNp_{eff}}{2} = \frac{zNp}{2} + n_c^d N_d, 
\end{equation}
where the equivalent passive network has the same number of constraints as that of our simulated network with dipoles. Here, $p_{eff}$ should be interpreted as the effective bond probability $p$ of the actively pre-stressed network with dipoles, which captures the effect of additional constraints imposed by the dipoles. Thus, we expect the active network with bond probability $p$ to have the same macroscopic mechanical properties, e.g., shear modulus, on average, as an unstressed, passive network (without dipoles) with bond probability $p_{eff}$. Using $z=6$ for 2D triangular lattice and re-expressing this relation as,
\begin{equation} \label{eq:peff}
    p_{eff} = p + \frac{n_c^d}{3N}N_d
\end{equation}
we can estimate, using the fitted slope of the linear data in Fig. ~\ref{fig:peff} $dp_{eff}/dN_d = 2.44 \times 10^{-3}$ and the total number of network nodes, $N=2347$, that the effective number of constraints per dipole is $n_c^d = 17.2 \pm 0.1$. We also show through the analysis in Appendix A and SI, Fig S14 that this value is practically unchanged for a network that is twice as large, and is thus quite robust to system size. The results in Fig. 4 are for bending-dominated networks, which is a deliberate choice to compare networks in the same regime, and to avoid effects due to bend-to-stretch phase transition. The  corresponding results for all networks, including those that turn stretching-dominated, are presented in the SI, Fig. S15.
Interestingly, we see in Fig. ~\ref{fig:peff}b  that the  effective bond probability, $p_{eff}$, values are nearly independent of the fiber bending modulus, $\widetilde\kappa$, within error bars. In contrast, Fig. ~\ref{fig:peff}a shows that network stiffness and consequently, the boundary forces measured by $\langle D_{far} \rangle$, scale with $\widetilde\kappa$. This is because $p_{eff}$ is directly related to the number of floppy bending modes available for the dipoles to remove, and not to the energy stored in these modes which scales with $\widetilde\kappa$. 
\\ \\
A more sophisticated counting argument is required to justify why $n_c^d  \simeq 18$ for this model of dipoles, and how this depends on the bond probability. A naive count suggests that each isotropic dipole imposes a rest length change on six surrounding bonds, and thus $n_c^{d} = 6$. However, elastic force transmission is non-local, and the dipole-imposed forces travel beyond the six immediate dipole bonds whose rest length is reduced.  The forces that extend beyond the immediate vicinity of the dipole and affect other nodes in the network possibly imposing  partial (fractional) constraints, the number of which decays with distance from the dipole.  Thus, we expect $n_c^d = 6$ to be a lower bound on the number of constraints per dipole, that is possibly realized when the dipoles are packed closer together. We also expect the number of constraints per dipole to depend on the network coordination.  For a network close to the isostatic point of $p \simeq 0.67$, we expect that there are fewer soft bending modes that the dipoles can remove, and thus less additional constraints each dipole adds.  In this work, we treat $n_c^{d}$ as a fitting parameter obtained from our simulation results, that provides insight into the constraints added per dipole.  The question of how the number of constraints per dipole depends on the bond probability as well as the ratio of bending to stretching moduli, is left as a topic of future investigation.

\subsection{Local architecture of force dipole significantly affects macroscopic network mechanical response}

We now investigate how the results presented thus far, including the approximate number of constraints per dipole, depend on the specific way in which these local dipole deformations are applied. As illustrated in Fig. ~\ref{fig:three_models})a-c, we consider three specific ways in which the local active units are coordinated. In all three cases, the six radial bonds emanating from the central node of the dipole are present, and their rest length reduced to generate contractile stress. In Model 1 considered so far, all six transverse bonds that connect the outer dipole nodes are removed.  We now introduce Model 2, where these transverse bonds are allowed to be randomly present according to the overall network bond probability ($p$) value, and Model 3, where all six transverse bonds are retained. We thus expect the extent of constraints to progressively increase from Model 1 to 3.

\begin{figure*}[htp]
    \centering
    \includegraphics[width=18cm]{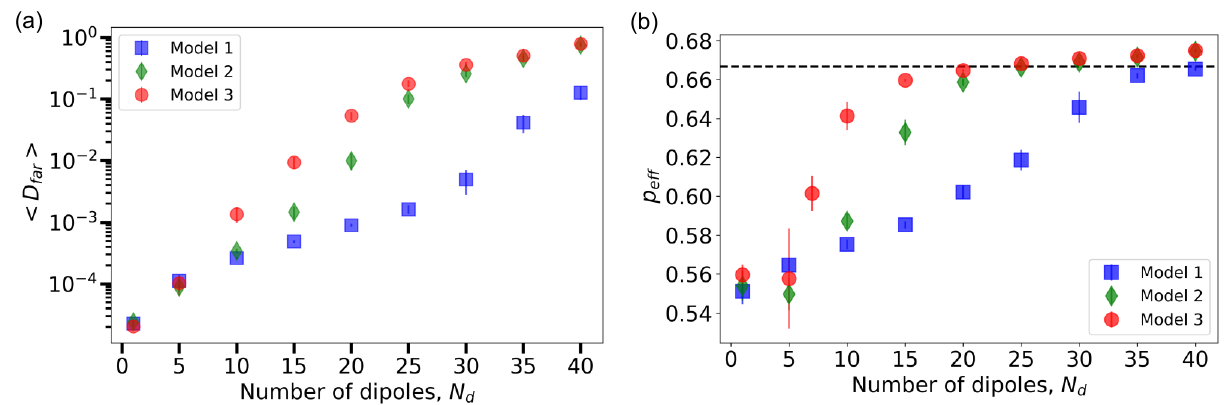}
    \caption{\textbf{Dependence of far-field dipole moment and effective connectivity on dipole number for the three local dipole models in all networks (both bending and stretching-dominated at $\boldsymbol{p=0.55}$ and $\boldsymbol{\widetilde\kappa=10^{-6}}$.}
    (a) The three models are seen to have similar $\langle D_{far} \rangle$ values at lower dipole numbers ($N_d$). They show separation by a few orders of magnitude at intermediate dipole numbers ($10 < N_d < 40$), before apparently converging at higher $N_d$.
    (b) Effective network connectivity, $p_{eff}$ values, obtained from EMT (see main text), show different rates of increase with dipole number for the three models. This value starts at the bare network value of $p_{eff} = p  =0.55$ for all three models at low dipole numbers ($N_d = 1,\, 5$), and saturates to the isostatic value, $p_{eff} = p_{cf}  =0.67$  at high $N_d$. 
    However, at intermediate dipole number ($5 < N_d < 30$), the three models give very different $p_{eff}$ values. The eventual saturation to the isostatic value may be understood as follows. 
   Once the network is stiffened to the isostatic point, there are no further soft bending modes left for additional dipoles to ``pull out''.
   Each plotted value of $\langle D_{far} \rangle$ is an average over $100$ simulations ($10$ dipole configurations $\times 10$ network configurations), except for the one dipole case, where  averaging is over 1000 simulations. Error bars indicate 95 $\%$ confidence intervals on the mean value obtained from bootstrapping.
    }
    \label{fig:three_models_dfar}
\end{figure*}

We show in Fig.~\ref{fig:three_models} that the specific manner of application of a localized isotropic force dipole has a significant impact on network deformation. Figs.~\ref{fig:three_models} a, b, and c, correspond to Models 1, 2 and 3, respectively. In each case, we show a representative network simulation snapshot, with five dipoles each ($N_{d}=5$). The network in \ref{fig:three_models} c (Model 3) has more visible ``force chains'' (shown in color) than \ref{fig:three_models} b and a (Models 2 and 1). In Figs.~\ref{fig:three_models} d, e, and f, we plot the distribution of ratios of stretching and bending energy values for all $1000$ network configurations simulated with $N_d = 5$ dipoles. This ratio indicates whether the network elastic energy is bending- or stretching-dominated. The transition from bending to stretching-dominated is known to occur with increasing network coordination number \cite{das2012redundancy, broedersz2011criticality}, or with applied external shear \cite{sharma2016strain, shivers2019scaling, arzash2021shear}. It is characterized by a sharp increase in network stiffness and is related to the nonlinearity of the boundary stress, $D_{far}$, with increasing dipole number.

As shown in Figs.~\ref{fig:three_models} d,e,f, we find that for $N_{d} =5$, all simulations for Model 1 result in bending-dominated ($E_{bend}/E_{stretch} > 1$) configurations, while for Model 2 and Model 3, a significant fraction of the simulated networks turn stretching-dominated ($E_{bend}/E_{stretch} < 1$). The number of stretching-dominated networks increases from Model 2 to Model 3, with increasing local coordination of the active units. Altogether, we show that even for the same overall network and dipole configuration, increasing local dipole constraints can drive the networks from the bending to the stretching-dominated regime.

We now compare the dependence of the boundary stress, $\langle D_{far} \rangle$, on dipole number, for the three models in Fig. ~\ref{fig:three_models_dfar}. As Fig.~\ref{fig:three_models} shows, models 2 and 3 result in a large proportion of stretching-dominated networks. Therefore, for effective comparison between the models, we now include all networks, including those that are stretching-dominated, in our analysis. Thus, we depart from  previous Figs. ~\ref{fig:dfar_nd}- ~\ref{fig:peff}, where  $\langle D_{far} \rangle$ and $p_{eff}$ were reported by averaging over only the bending-dominated networks. At low dipole numbers ($N_d < 10$), the average boundary stress $\langle D_{far} \rangle$ starts out similar for the three models. The $\langle D_{far} \rangle$ values for Model 3, which is locally more coordinated at the dipoles, begin to separate from Models 1 and 2 at $N_d = 10$. Subsequently, Model 2 values separate from Model 1 at $N_d = 15$, indicating the differences in onset of the stiffening transition in the different models. As $N_d$ increases further, eventually the three model results converge.

These observations may be rationalized by the differences in the way the three models impose local constraints.  To demonstrate this, we carry out for all models the same analysis for $p_{eff}$ using EMT, that was done for Model 1 in Fig.~\ref{fig:peff}. The results, shown in Fig. ~\ref{fig:three_models_dfar}b, correspond to the $\langle D_{far} \rangle$ plots of Fig. ~\ref{fig:three_models_dfar}a.  The intermediate result for the ratio of far field dipole moments at $p=0.55$ and $p=1$ ,which lets us determine $\mu_m$, is shown in the SI, Fig. S16. We demonstrate that the effective coordination numbers, induced by the dipoles, are significantly different for the three different dipole models and progressively increase from Model 1 to 3. 

We see in Fig. ~\ref{fig:three_models_dfar}b that with increasing dipole density,  all three models saturate near a $p_{eff} \simeq 0.67$. This is in fact the isostatic point for a 2D triangular spring network, $p_{CF}$. Since force dipoles ``pull out'' floppy modes to stiffen the network, the stiffening effect is  expected to saturate when the network effectively reaches the isostatic point. Beyond this threshold value, there are no more floppy modes left to pull out.  The addition of more dipoles does not  further stiffen the network appreciably. The different onsets of the stiffening transition with number of dipoles in Fig. ~\ref{fig:three_models_dfar}a becomes more apparent in Fig. ~\ref{fig:three_models_dfar}b, as the differences in the onset of saturation to $p_{eff} = p_{CF}$.  We note that while we estimated the effective number of constraints imposed by each dipole for bending-dominated networks in Model 1 to be $n^{d}_{c} \simeq 17$ from the linear regime of the data in Fig. ~\ref{fig:peff}, the actual data is nonlinear. The additional constraints also eventually saturate for Model 1 at higher dipole density. On the other hand, we see that all three models exhibit a linear increase in $p_{eff}$ vs $N_d$ in Fig. ~\ref{fig:three_models_dfar}b at intermediate dipole number. From the corresponding slopes, we can estimate the number of constraints per dipole, $n^{d}_{c}$, which progressively increases from Models 1 to 2 to 3, as expected. Overall, all three models show qualitatively similar stiffening with increasing dipole number, while differing in the specific number of constraints imposed.
        
\section{Discussion}

In this work, we quantitatively explored the transmission of actively generated mechanical forces in a disordered fiber network. We developed a model featuring contractile, isotropic force dipoles embedded in bond-diluted triangular lattice of springs with bending and stretching, which can represent biologically relevant systems such as the cytoskeleton or cells in extra-cellular matrix. By numerically minimizing the elastic energy of many realizations of such elastic networks, we show how macroscopic network contractility emerges for a collection of such force dipoles that mutually deform and stiffen the network.

A main result of our work is the nonlinear scaling of the ``far-field dipole moment'', a measure of network contractility, with increasing force dipole density. This apparent stiffening of the network under internal pre-stress is consistent with previous theoretical works \cite{broedersz_11, Chen2011, Sheinman2012ActivelyStressed} and experiments on biopolymer networks \cite{koenderink2009active, sharma2016strain}. Unlike these previous works, which directly measure the network shear modulus, we demonstrate the stiffening transition by measuring the isotropic stress at network boundary \cite{Ronceray16}, a quantity directly related to network contractility. We showed that this boundary stress scales non-linearly with dipole number in Fig. \ref{fig:dfar_nd} for under-coordinated ($p < p_{CF}$) and bending-dominated networks. The non-linear scaling is independent of fiber bending modulus, as seen in Fig. \ref{fig:dfar_nd}b. This result implies that a pre-stressed network becomes more efficient at transmitting forces since individual fibers undergo more stretching than bending, once the floppy modes of the network are ``pulled out'' by the dipoles. On the other hand, we showed that boundary stress scales linearly with bending modulus, consistent with the linear scaling of network shear modulus with bending modulus in the bending-dominated regime \cite{head2003distinct}. We also found in Fig.~\ref{fig:dfar_en} that the boundary force scales differently with bending and stretching energies of the network, showing how different deformation modes compete and transmit force differently. 

While it has long been appreciated that force dipoles stiffen elastic networks by imposing constraints on floppy modes \cite{broedersz_14}, we sought to quantitatively answer the question: how many additional constraints does a force dipole impose on the network? We utilize the attenuation of the boundary stress by bond dilution to infer an effective stiffness of the network, $\mu_m$. We then compare with established effective medium theory (EMT) to extract an effective coordination number for the dipole-stressed network.  This apparent coordination, $p_{eff}$, is higher than the physical coordination, $p$, of the network. The enhanced effective connectivity by dipoles is independent of fiber bending modulus, since it depends on the number, and not energy, of floppy bending modes removed by the dipoles. In fact, for an intermediate regime of dipole density, we find that $p_{eff}$ increases linearly with increasing dipole number, corresponding to a fixed number of additional constraints imposed by each dipole. We find that for our specific first choice of the dipole model, this count is about a factor of three higher than the naive lower bound estimate of $6$ constraints per dipole, corresponding to the six springs attached to the central dipole node.  We speculate that this enhanced count is due to the dipole-imposed stress spreading through the elastic network, thereby creating additional constraints on network nodes not in the immediate vicinity of the force dipole.

We note that the active stresses in this model are spatially localized, internal force dipoles that build pre-stress in the network. We do not consider other mechano-chemical adaptation and feedback effects that require energy consumption and are crucial in living matter. Adherent cells, for example, can regulate their contractility in response to deformations in the extra-cellular matrix \cite{zemel_10, SirotePRE2021}. Interestingly, even without these feedback effects that also occur in principle in linear elastic media \cite{schwarz_02}, the inherent nonlinearity of disordered elastic networks leads to cooperative effects between these force dipoles, manifest as increased boundary stress or contractility per dipole, at higher dipole density. These effects are particularly prominent in the sub-isostatic, bending-dominated regime. This has been shown to be the relevant regime for biopolymer networks, allowing them to strongly strain-stiffen \cite{sharma2016strain}.  Our result for reinforced contractility through stiffening may therefore relate to experimental observations of a sharp increase in collagen gel contraction by fibroblasts above a threshold cell density \cite{doha2022disorder}.

Finally, we explored different architectures of the force dipole and their impact on the boundary stress, $\langle D_{far} \rangle $. We considered three different local coordinations of the active units that impose dipole forces, and found that this has large consequences for the emergent macroscopic contractility. Specifically, the strength and onset of stiffening, and correspondingly, the number of effective constraints imposed by force dipoles, vary sensitively between these three models. 
 
This sensitivity to dipole model is clearly seen in Fig. ~\ref{fig:three_models_dfar}a at intermediate values of the dipole density ($15 \le N_d \le 35 $).  In this regime, the boundary stress  values, given by $\langle D_{far} \rangle$, can differ between the three models by one to two orders of magnitude. 
This reveals an optimal range of dipole density where the mechanical response of elastic networks is maximally sensitive to active stress. Within this range, small changes in local dipole constraints can result in large changes in macroscopic network mechanical response. This sensitivity may provide biological systems, such as myosin molecular motors in the cell cytoskeleton, and cells adhered to an extra-cellular matrix, an efficient strategy to regulate stiffness and force transmission, simply by making local structural changes to their surrounding elastic network. 

It is well-known that myosin motor activity tunes cytoskeletal stiffness \cite{koenderink2009active} and network contractility \cite{alvarado2013molecular}, and that collagen networks are highly sensitive to applied stress \cite{sharma2016strain}. Our work suggests that cells may further regulate the onset and extent of this stiffening by tuning how they apply forces to the surrounding network. Cells such as fibroblasts may tune the local coordination of their network by degrading or depositing collagen fibers in their vicinity \cite{Wershof2019MatrixFeedback}, whereas myosin molecular motors may do so by locally cross-linking or fracturing actin filaments  \cite{backouche2006active, haviv2008cytoskeletal}. This in turn may tune between local and global contractile regimes of the network, seen in the cytoskeleton \cite{AlvaradoSM2017, norman2025connectivity}.

Thus,  our observed stiffening of fiber networks by active forces, and the sensitivity of this effect to  the local structural details of the force-producing units, has practical mechanobiological implications for cell mechano-sensing and response.

\section*{Author contributions}
AK performed the simulations and analyses. All authors contributed to the conceptualization, discussion of results and writing of the final manuscript.

\section*{Acknowledgements}

AK and KD acknowledge support from the National Science Foundation through CAREER award DMR-2340632 to KD. AK and KD also acknowledge computational resources through the NSF CREST: Center for Cellular and Biomolecular Machines (CCBM) at the University of California, Merced through grant HRD-1547848. KD also acknowledges the Aspen Center for Physics, which is supported by National Science Foundation grant PHY-2210452, where part of this work was performed. 
DAQ acknowledges that this work was performed under the auspices of the U.S. Department of Energy by Lawrence Livermore National Laboratory under Contract DE-AC52-07NA27344 (LLNL-JRNL-2011080). 

\section*{Appendix}

\appendix
\renewcommand{\thefigure}{\Alph{section}\arabic{figure}}
\renewcommand{\theequation}{\Alph{section}\arabic{equation}}

\section{System size effects}
\setcounter{figure}{0}
\setcounter{equation}{0}

\begin{figure*}[!ht]
    \centering
    \includegraphics[width=15cm]{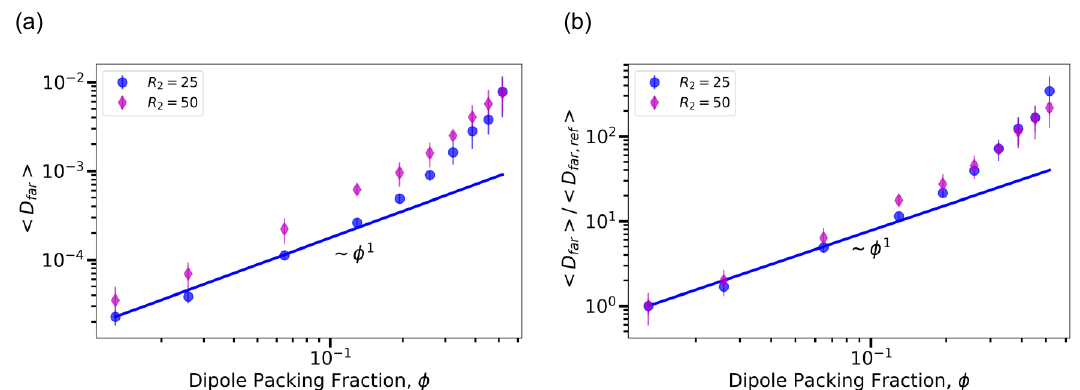}
    \caption{ \textbf{$\boldsymbol{D_{far}}$ scales non-linearly with dipole number independently of system size.}
    (a) Far field dipole moment of a network with twice the radius of the original network agrees well with that of the original network. Furthermore, the large network also indicates a non-linear increase in $D_{far}$ with increasing number of dipoles. Here, the packing fraction represents the area density of dipoles where each dipole occupies 5.84567 area which is the area of a regular hexagon with side length of 1.5. The side length of 1.5 is used instead of 1 because the hexagons are non-overlapping. Therefore the packing fraction scales linearly with $N_d$.  
    (b) We plot the normalized far field dipole moment versus packing fraction to find that the normalized values of $D_{far}$ for $R_2=25$ and $R_2=50$ collapse onto each other. The normalization was done using the $D_{far}$ for $N_d = 1$ case for the smaller network and using $N_d = 4$ case for the larger network.
    }
    \label{fig:dfar_L128}
\end{figure*}

To test how robust the stiffening behavior is to system size variations, we simulated a larger network with $R_1 = 24$ and $R_2=50$. To compare the two different-sized networks, we first compute a packing fraction. We enforce that any two dipoles must be non-overlapping, therefore, each dipole has an area equivalent to a small regular hexagon of side 1.5 (total area per dipole = $\tfrac{3\sqrt{3}(1.5)^2}{2}$). The length $1.5$ is chosen such that two neighboring dipoles do not overlap and that they completely pack the area between them. The packing fraction is then the total area occupied by $N_d$ dipoles divided by the total area of the inner region ($\pi R_1^2$).

First, we quantify the scaling of $\langle D_{far} \rangle$ with the packing fraction to find that at high packing fractions (or high dipole numbers), the increase in the far-field dipole moment is 
non-linear with respect to increasing 

packing fraction in both systems (Fig. ~\ref{fig:dfar_L128}a) and that the $D_{far}$ values agree well between the two system sizes (Fig. ~\ref{fig:dfar_L128}a,b).For the large system size where $R_2 = 50$, SI, Fig. S12 shows that $p_{eff}$ also increases linearly with the number of dipoles just as in Fig. ~\ref{fig:peff}b. Repeating the constraint count calculation in the main text, now for the larger system size which has: $dp_{eff}/dN_d = 6 \times 10^{-4}$ and $N = 9271$, yields $n_c^d = 16.7 \pm 0.5$. Therefore, the number of constraints applied per dipole is robust to system size effects.

\section{Simulation methodology}
\setcounter{figure}{0}
\setcounter{equation}{0}

Here, we provide more detail on how the simulation setup is prepared. We set up a $L \times L$ lattice of equilateral triangles each of edge length unity. By measuring node distances from the the geometric center of this rectangular region, we define the outer boundary as the set of nodes whose positions satisfy: $R_2 - dr < r_{i} < R_2 + dr$, where $dr=0.1$.  We further ensure that each node on the boundary is connected to its neighboring boundary nodes to create a continuously connected boundary.  The force dipoles are randomly seeded in an inner region of $r \le R_1$. 

When preparing the network, we remove bonds in the lattice based on a pseudo-random number generator. If the randomly generated number (between $0$ and $1$) is below (above) the target bond occupancy $p$, the bond is retained (removed). 
 
We remove all dangling bonds from the network as they cost zero energy to rotate and therefore have no effect on the mechanical properties of the network. When we design the isotropic dipoles, all of six radial bonds are present at each central node of each dipole, in all three dipole models considered. These perturbations introduced at the force dipoles can change the average bond occupancy, $p$, of the network. However, we checked that the net change in the value of $p$ is sufficiently small ($1\% -2 \%$).  In the simulation, we  change the rest-length of the radial bonds of the dipole in $10$ steps. At each step, we minimize the network energy using the conjugate gradient method.

We performed a numerical convergence study with respect to the number of incremental relaxation steps used to update the radial-bond rest lengths. Increasing the number of steps (i.e., decreasing the per-step update magnitude) produced no statistically significant change in the post-relaxation network energies, after applying a conjugate-gradient minimization at each increment. This indicates that the converged energy minimum is insensitive to the specific relaxation discretization used for the dipole-bond updates, and is therefore robust with respect to the bond-relaxation protocol.

The stretching and bending forces are computed in the standard way \cite{monasse2014forces} by taking gradient of the energy expression in Eq.~\ref{eq:total_energy} . Here, we briefly state these expressions and provide intuitive arguments to justify the bending force without lengthy derivation. The stretching force on the $i^{th}$ node from the bond spring connecting it to the $j^{th}$ node, if present, is proportional to bond strain and is directed along the bond vector,
\begin{equation}
\mathbf{F}^{\langle ij \rangle}_{s,i} = -\mu (r_{ij} - r_{0}) \frac{\mathbf{r}_{ij}}{r_{ij}}   
\label{eq:stretching_force_ch3}
\end{equation}
where $\mathbf{r}_{ij}$ is the bond vector pointing from node $j$ to $i$, and $r_{ij}$ is its magnitude. 

We now consider the three-body bending forces generated by change of the angle $\theta_{jik}$, between initially co-linear bonds $ji$ and $ik$. Note that the bending force is generated only if both these bonds are present. The bending force from distortions of this angle on the peripheral $j^{th}$ node is obtained by taking gradient of the bending energy cost with respect to the position of this node. This contribution may be written as,

\begin{eqnarray}
    \mathbf{F}^{\langle jik \rangle}_{b,j} &=& \frac{ \kappa}{r_{0}} \sin \theta_{jik}
    \frac{\partial \theta_{jik}}{\partial \mathbf{r}_{j}} \\
    &=& -\frac{\kappa}{r_{0}} \sin \theta_{jik} \frac{\hat{z} \times \mathbf{r}_{ij}}{r^{2}_{ij}},   
 \label{eq:bending_force_1_ch3}   
\end{eqnarray}
where the cross product with the unit normal $\hat{z}$ to the plane of the three nodes, results in a vector that is perpendicular to the bond vector $\mathbf{r}_{ij}$. Thus, this expression captures the intuitive expectation that the gradient of the angle points in the direction where the angle maximally changes, that is, in the direction orthogonal to the bond vector. 
On similar grounds, the corresponding force on the other peripheral node $k$ may be written as,
\begin{equation}
     \mathbf{F}^{\langle jik \rangle}_{b,k} = -\frac{\kappa}{r_{0}} \sin \theta_{jik} \frac{\hat{z} \times \mathbf{r}_{ik}}{r^{2}_{ik}},
     \label{eq:bending_force_2_ch3}  
\end{equation}
whereas the force on the central node $i$ is just equal and opposite to the sum of the forces on the two other nodes, 
\begin{equation}
    \mathbf{F}^{\langle jik \rangle}_{b,i} = -(  \mathbf{F}^{\langle jik \rangle}_{b,j} + \mathbf{F}^{\langle jik \rangle}_{b,k}),
    \label{eq:bending_force_3_ch3}  
\end{equation}
which is required for internal force balance of the three body system. Note that we give here the expression for forces resulting only from the angular spring at node $i$. The net bending force on a given node can include up to three contributions: one where it is the central node, and two others with it being a peripheral node in the three-node system. 

\section{Measurement of local force dipole moment to quantify active forces: $D_{loc}$}
\setcounter{figure}{0}
\setcounter{equation}{0}

\begin{figure}[ht]
    \centering
    \includegraphics[width=6cm]
    {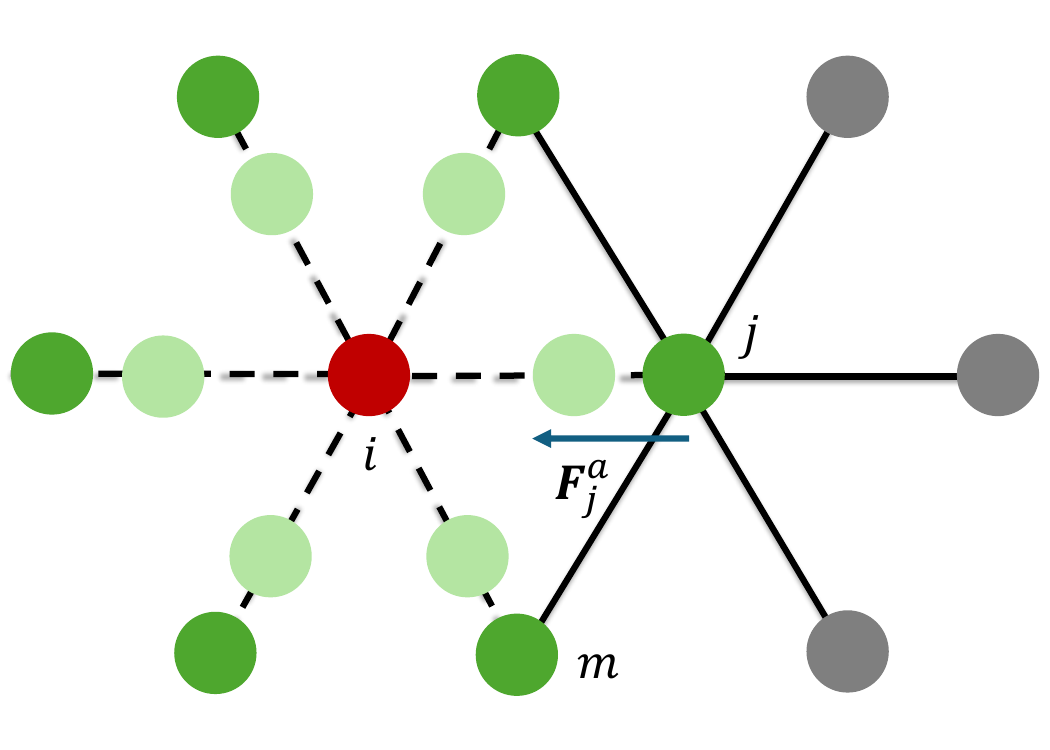}
    \caption{\textbf{Schematic illustrating calculation of $\boldsymbol{D_{loc}}$ for a single active unit.} For a given active unit producing an isotropic force dipole, the central dipole node is marked in red (node $i$). The six outer dipole nodes neighboring the central node, are in green (initial, undeformed positions) and light green (final, deformed positions), while the next nearest neighbor nodes  are in gray. The rest-length of all $6$ radial bonds of the central dipole node is set to $\bar{r}_{dip} = 0.9$, while the initial bond lengths are set to $1$, just as in the rest of the network. When the network relaxes towards force balance, the outer dipole nodes move inward, closer to the central dipole node. This inward displacement can equivalently be generated by an applied ``active force'' acting on each of the outer dipole nodes, denoted as $\mathbf{F}_j^a$ acting on the $j^{th}$ node. The local dipole moment due to this active unit, $D_{loc}$, is the sum of moments of each of these active forces about the $i^{th}$ node, see Eq.~\ref{eq:dloc_i}.
    Force balance at the $j^{th}$ node implies that $\mathbf{F}_j^a$ is the opposite of the sum of the spring forces (both bending and stretching) acting on the $j^{th}$ node, due to all its  neighboring  $m^{th}$ nodes, see Eq.~\ref{eq:active_force}.
    }
    \label{fig:dloc-schematic}
\end{figure}

We use the local dipole moment of active forces, labeled $D_{loc}$, as a measure of the amount of active stress exerted by the actively contractile units \cite{Ronceray16, ronceray2015connecting}.  This process for calculating the local force dipole moment is illustrated in figure \ref{fig:dloc-schematic}.  Essentially, we calculate the active force that would need to be applied to each outer dipole node (labeled by index $j$) to reach the same deformed configuration that is attained by reducing the rest length of the dipole bonds.
Say we are at the $j^{th}$ outer node around the $i^{th}$ central dipole node, connected by the bond vector $\mathbf{r}_{ij}$. We calculate the active force acting on the $j^{th}$ node from the force-balanced configuration. The scalar product of this force with the separation vector joining this node to the central dipole node gives the contribution to the local dipole moment from this node:
\begin{equation} \label{eq:dloc_i}
D_{loc, j} = \sum_{m} \mathbf{F}_{j}^a \cdot \mathbf{r}_{ij}  
\end{equation}

Using force balance, the active force $\mathbf{F}_j^a$ is the opposite of all the stretching and bending forces acting on the $j^{th}$ node, that is, 
\begin{equation} \label{eq:active_force}
F_j^a = - \sum_{m} \mathbf{f}_{mj} - \mathbf{F}_{b,j}, 
\end{equation} 
where $\mathbf{f}_{mj}$ is the stretching force of each bond connecting the $j^{th}$ node to its neighboring $m^{th}$ node.  
$\mathbf{F}_{b,j}$ is the total bending force acting on the $j^{th}$ node, due to all possible connected, collinear node triplets involving the $j^{th}$, see Appendix B. %Eq. ~\ref{eq:bending_force_1_ch3}.
The stretching force in each bond connecting $j^{th}$ to  $m^{th}$ node, when present, is given by, 
$\mathbf{f}_{mj} = \mu (|r_{mj}| - r_0) \mathbf{\hat{r}}_{ij}$,
where $r_{0} =1$ is the initial bond length in the undeformed triangular lattice, and $r_{mj}$ is the corresponding deformed bond length.
Importantly, we don't consider the modified rest length change in this expression. This is because the quantity of interest we want to calculate here is the equivalent active force which would produce the same deformed configuration as the rest length change. If a node is fully coordinated, there can be a maximum of 3 pairs of collinear bonds that pass through it, each contributing to the value of total bending force on the node. The node can have bending force contributions from configurations where it is not the central node, but is the periphery node of a three-node system that defines a pair of collinear bonds. The total bending force $\mathbf{F}_{b,j}$ includes contributions from all such possible combinations.

Then, summing over all six outer dipole nodes ($j^{th}$ nodes) for each central node ($i^{th}$ node) of a force dipole and repeating the process for each force dipole gives us the total local dipole moment: $D_{loc} = \sum_{i} \sum_{j} D_{loc,j}$.

\section{Mean Stress theorem and Dipole conservation}
\setcounter{figure}{0}
\setcounter{equation}{0}

Here, we show that for an elastic body undergoing deformations,  the mean stress can be related to the difference of the boundary and local dipole moment tensors. 

Consider a 2D elastic body of arbitrary shape. It may be represented by a 2D domain $\Omega$ bounded by curve $\delta \Omega$. When acted by an active force density, $\mathbf{f}^{a}$, the body deforms and develops an elastic stress that satisfies force balance,
\begin{equation}
    \partial_{\gamma} \sigma_{\alpha \gamma} = f^{a}_{\alpha}, 
 \label{force_bal}   
\end{equation}
where Greek indices denote spatial coordinates, as opposed to Latin indices used to denote discrete nodes in the simulation model. We also use usual Einstein summation convention, where repeated indices imply summation.

We now relate this local force balance condition to macroscopic stresses measured at the boundary.  As a first step, we take the moment of the forces on both sides of Eq.~\ref{force_bal}  and integrate over the entire domain to obtain,
\begin{equation}
    \int_{\Omega} dA \, r_{\beta}  \, \partial_{\gamma} \sigma_{\alpha \gamma}   = -  \int_{\Omega} dA \, r_{\beta} f^{a}_{\alpha}  
\label{force_bal_moment}    
\end{equation}
Integrating by parts, and using the divergence theorem in 2D (also known as the Green-Gauss theorem), the left hand side of Eq.~\ref{force_bal_moment} can be re-expressed as:
\begin{equation}
    \int_{\Omega} dA \, r_{\beta}  \partial_{\gamma} \sigma_{\alpha \gamma}  =
    \int_{d\Omega} dl \, n_{\gamma} r_{\beta} \sigma_{\alpha \gamma}  - \int_{\Omega} dA \, \sigma_{\alpha \beta},  
\label{integral_moment}    
\end{equation}
where the first term represents a flux over line element $dl$ on the closed boundary of the elastic domain. We now use the elastic boundary condition, that is the definition of the stress tensor on the bounding surface,  $ \sigma_{\alpha \gamma} n_{\gamma} = f^{b}_{\alpha} $, where ${\bf f}^{b}$ is the force (per unit length) on the bounding surface. Using this re-expressed form of Eq.~\ref{integral_moment} in the integrated moment balance of Eq.~\ref{force_bal_moment}, we obtain a statement of the \emph{mean stress} theorem:
\begin{equation}
    D^{far}_{\alpha \beta} - A \bar{\sigma}_{\alpha \beta}  = D^{loc}_{\alpha \beta}   
\label{mean_stress}    
\end{equation}
where the boundary dipole moment is given by
$ D^{far}_{\alpha \beta} = \int_{d\Omega} dl \, r_{\beta} f^{b}_{\alpha}$,
the integrated dipole moment of the active forces (localized, in all cases we consider) is $ D^{loc}_{\alpha \beta} = \int_{\Omega} dA r_{\beta} f^{a}_{\alpha}$, and the mean stress is just the stress integrated over the whole domain divided by its area, $A$.
Note that this result is obtained from force balance. Hence, it is general and does not require any specific constitutive relation. It holds for an inhomogeneous and nonlinear elastic medium, but is of limited applicability because the mean stress in the second term needs to be computed over the whole domain.  

We now show that for a linear elastic medium with clamped boundary conditions, the mean stress vanishes.  
The stress tensor in 2D linear elasticity is expressed in terms of shear and bulk strain as,
\begin{equation}
   \sigma_{\alpha \beta} = 2 \mu u_{\alpha \beta} + \lambda u_{\gamma \gamma} \delta_{\alpha \beta},   
\label{linear_elasticity}   
\end{equation}
where $u_{\alpha \beta} = 1/2 (\partial_{\alpha} u_{\beta} + \partial_{\beta} u_{\alpha}) $ is the symmetric linear strain tensor obtained as gradient of small displacement of material points, and $\lambda$ and $\mu$ are the Lame moduli \cite{landau_lifshitz_elasticity}.

Integrating Eq.~\ref{linear_elasticity} over the whole domain and applying the divergence theorem, we get,

\begin{align}
    \int_{\Omega} dA \, \sigma_{\alpha \beta} 
    &= \mu \int_{\Omega} dA \, (\partial_{\alpha} u_{\beta} + \partial_{\beta} u_{\alpha}) 
    + \lambda \delta_{\alpha \beta} \int_{\Omega} dA \, \partial_{\gamma} u_\gamma \nonumber \\
    &= \mu \int_{\partial\Omega} dl \, (n_{\alpha} u^{b}_{\beta} + n_{\beta} u^{b}_{\alpha}) 
    + \lambda \delta_{\alpha \beta} \int_{\partial\Omega} dl \, n_{\gamma} u^{b}_{\gamma},
\end{align}
where ${\bf n}$ is the unit normal to the boundary and ${\bf u}^{b}$ is the displacement at the boundary. Therefore, for clamped boundaries, ${\bf u}^{b} =0$, and the integrated stress or mean stress disappears, $\bar{\sigma}_{\alpha \beta} =0$. Using this in the mean stress theorem obtained in Eq. ~\ref{mean_stress} implies the conservation of dipole moments, ${\bf D}_{far} = {\bf D}_{loc}$. 

This motivates the calculation of the traces of these dipole moment tensors ${\bf D}_{far}$ and ${\bf D}_{loc}$ in the main text. For $p=1$ networks, our simulations shown in SI, Fig. S4, verify
$D_{far} \simeq D_{loc}$. This is in accordance with the result just derived, because the fully coordinated $p=1$ network corresponding to triangular lattice is linearly elastic under small deformations. However, for $p< 0.67$, i.e. sub-isostatic or under-constrained networks, we do not expect this relation to be true. In fact, there we find that $D_{far} \ll D_{loc}$. Intuitively, this is because the force travels through a small number of springs. Thus, we use the quantity $D_{far}/D_{loc} < 1$ as a measure of the amount of applied force that is transmitted to the boundary.

\typeout{}
\bibliography{bibliography}
\bibliographystyle{rsc}

\end{document}